\documentclass[aps, prb, amsmath, amssymb, reprint, superscriptaddress,]{revtex4-1}
\setcitestyle{super}
\usepackage{graphicx}
\usepackage{dcolumn}
\usepackage{units} 
\usepackage{bm}
\usepackage{textgreek}
\usepackage{color,soul}
\usepackage{comment}
\usepackage{multirow}
\usepackage{lipsum}  
\usepackage{xcolor}
\usepackage{subscript}

\begin{document}

	\preprint{AIP/123-QED}
	
	\title{Asymmetric velocity and tilt angle of domain walls induced by spin-orbit torques}
	
	\author{Manuel Baumgartner}
	\affiliation{Department of Materials, ETH Zurich, H\"onggerbergring 64, CH-8093 Zurich, Switzerland}
	
	\author{Pietro Gambardella}
	\affiliation{Department of Materials, ETH Zurich, H\"onggerbergring 64, CH-8093 Zurich, Switzerland}
	
	\date{\today}
	
\begin{abstract}
We present a micromagnetic study of the current-induced domain wall motion in perpendicularly magnetized Pt/Co/AlO\textsubscript{x} racetracks. We show that the domain wall velocity depends critically on the tilt angle of the wall relative to the current direction, which is determined by the combined action of the Dzyaloshinskii-Moriya interaction, damping-like, and field-like spin-orbit torques. The asymmetry of the domain wall velocity can be controlled by applying a bias-field perpendicular to the current direction as well as by the current amplitude. As the faster domain walls are expelled rapidly from the racetrack boundaries, we argue that the domain wall velocity and tilt measured experimentally depend on the timescale of the observations. Our findings reconcile the discrepancy between time-resolved and quasi-static domain wall measurements n which domain walls with opposite tilts were observed and are relevant to tune the velocity of domain walls in racetrack structures.
\end{abstract}
\maketitle

The propagation of domain walls (DWs) plays a fundamental role in determining the efficiency and speed of current-induced switching of magnetic devices.\cite{Grollier:2003aa, Klaeui:2005aa, Parkin:2008aa, Miron:2011ab, Emori:2013aa, Ryu:2013aa, Haazen:2013aa, Yang:2015aa, Safeer:2016aa, Baumgartner:2017aa} In the context of spin-orbit torques (SOTs),\cite{Manchon:2018aa} DW propagation has been extensively studied by analytical\cite{Thiaville:2012aa, Khvalkovskiy:2013aa, Martinez:2013aa, Boulle:2014aa} and micromagnetic models,\cite{Martinez:2014aa, Martinez:2013ab, Mikuszeit:2015aa, Lee:2018aa} magneto-optical Kerr effect (MOKE),\cite{Miron:2011ab, Emori:2013aa, Ryu:2013aa, Haazen:2013aa, Yang:2015aa, Safeer:2016aa} nitrogen-vacancy magnetometry,\cite{Tetienne:2014aa} as well as x-ray imaging.\cite{Baumgartner:2017aa, Vogel:2012aa} An important conclusion drawn from this extended body of work is that the DWs in perpendicular magnetized layers, such as Pt/Co/AlO\textsubscript{x} and Ta/CoFeB/MgO, are chiral N\'eel walls stabilized by the Dzyaloshinskii-Moriya interaction (DMI). The N\'eel wall magnetization points in-plane, perpendicular to the DW and hence parallel to the current direction, which maximizes the amplitude of the current-induced damping-like SOT and promotes very large DW displacement velocities $v_{\mathrm{\scriptscriptstyle{DW}}}$, of the order of $\unit[100]{m\ s^{-1}}$ for a current density $j=\unit[10^8]{A\ cm^{-2}}$. This large $v_{\mathrm{\scriptscriptstyle{DW}}}$ allows for high speed DW displacements in racetrack structures\cite{Miron:2011ab,Yang:2015aa,Safeer:2016aa} as well as for sub-ns reversal of ferromagnetic dots.\cite{Baumgartner:2017aa, Garello:2014aa}

\begin{figure}
	\centering
	\includegraphics[width=0.48\textwidth]{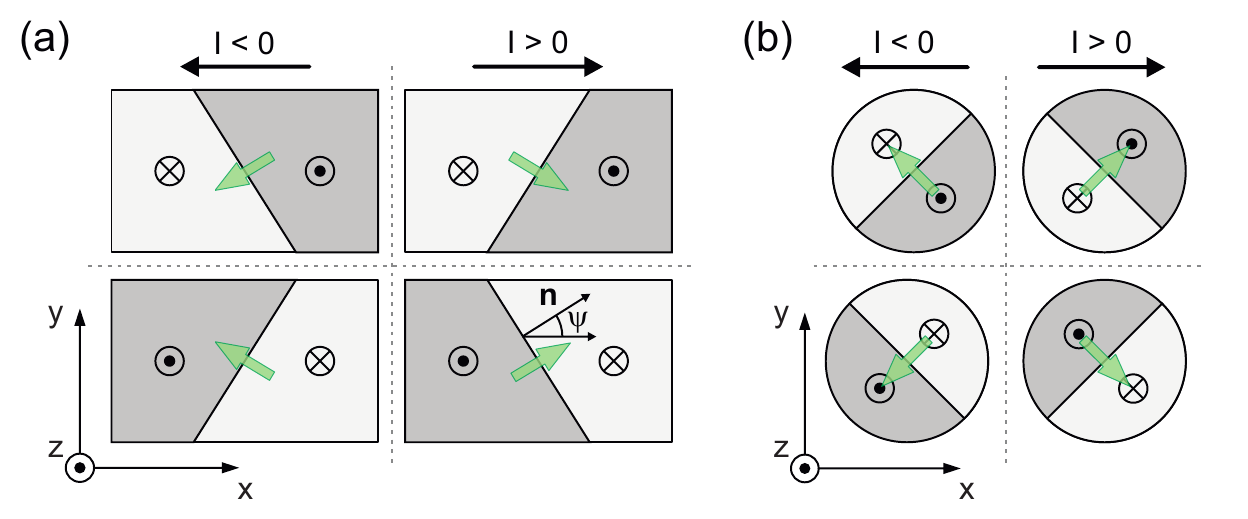}
	\caption{Schematics of the current-induced tilted DWs for the four combinations of current and domain orientation measured by (a) static MOKE microscopy.\cite{Ryu:2012aa} and (b) time-resolved x-ray microscopy\cite{Baumgartner:2017aa} in perpendicularly magnetized Pt/Co bilayers. The black and green arrows indicate the current and the propagation direction of the DWs, respectively. The tilt angle $\psi$ between the positive $x$-axis and the normal to the DW $\mathbf{n}$ is shown in (a).}
	\label{DW_tilt_symmetry:pic}
\end{figure}

Two prominent effects of the DMI in perpendicularly magnetized layers are the tilting of the DW\cite{Baumgartner:2017aa, Ryu:2012aa, LoConte:2017aa, Boulle:2013aa, Martinez:2016aa, Martinez:2014aa, Emori:2014ab} and the asymmetric $v_{\mathrm{\scriptscriptstyle{DW}}}$ relative to the current direction.\cite{Safeer:2016aa, Martinez:2016aa, Garg:2018aa} These two effects are related by the DW dynamics under the combined action of DMI and damping-like SOT.\cite{Boulle:2013aa, Martinez:2016aa, Martinez:2014aa, Emori:2014ab} Tilted DWs were first observed in Pt/Co/Ni/Co layers by imaging the magnetic domains after a sequence of current pulses using MOKE microscopy\cite{Ryu:2012aa} and later reproduced by analytical and micromagnetic models.\cite{Boulle:2013aa, Martinez:2014aa, Emori:2014ab} Figure~\ref{DW_tilt_symmetry:pic}(a) illustrates the DW configurations reported in Ref.~\onlinecite{Ryu:2012aa} for the four combinations of current (black arrows) and up/down, down/up domains propagating in a racetrack. The tilt angle is indicated by $\psi$ and the propagation direction of the DW is given by the green arrows. These DW tilt symmetries are typical of perpendicularly magnetized films with a Pt underlayer. Recent time-resolved x-ray microscopy measurements on Pt/Co/AlO\textsubscript{x} dots, however, reported DWs rotated by about $90{^\circ}$ for the same current polarity and domain orientation,\cite{Baumgartner:2017aa} as shown in Fig.~\ref{DW_tilt_symmetry:pic}(b). We suppose that these contrasting observations may arise from the static \textit{vs.} time-resolved nature of the experiments, since in the first case the DWs are imaged \emph{after} the injection of several current pulses, whereas in the second case the DWs are imaged \emph{during} current injection following a nucleation event. Furthermore, the field-like component of the SOT may also induce a tilt of the DW, similar to the effect of an in-plane field orthogonal to the current.\cite{Boulle:2013aa, Martinez:2014aa, Muratov:2017aa,Kim:2018aa} The aim of this work is to reconcile these controversial observations by elucidating the time-resolved dynamics of tilted DWs in racetrack structures and investigate the influence of DW tilt and field-like torque on the velocity of the walls.

We present a study of the current-driven dynamics of chiral DWs in heavy metal / ferromagnetic racetracks performed using micromagnetic simulations. As model system, we choose Pt/Co/AlO\textsubscript{x} stripes divided into $\unit[4]{nm} \times \unit[4]{nm} \times \unit[1]{nm}$ rectangular cells with the following material parameters: Co thickness \unit[1]{nm}, saturation magnetization $M_s = \unit[900]{kA\ m^{-1}}$, exchange coupling $A_{ex} = \unit[10^{-11}]{J\ m^{-1}}$, effective uniaxial anisotropy energy $K_u = \unit[657]{kJ\ m^{-3}}$, DMI constant $D = \unit[1.2]{mJ\ m^{-2}}$, and damping $\alpha = 0.5$. The magnitudes of the damping-like and field-like SOTs are given in field units per unitary magnetization as $T^{DL} = \unit[18]{mT}$ and $T^{FL} = \unit[10]{mT}$ per $j = \unit[1\times10^8]{A\ cm^{-2}}$, respectively. For simplicity, we neglect the effects of pinning and temperature,\cite{Martinez:2016aa, Martinez:2014aa} which are not central to the results presented in this work. The simulations were carried out using the object oriented micromagnetic framework (OOMMF) code\cite{Donahue:1999aa} including the DMI extension module\cite{Rohart:2013aa} as well as an additional SOT module. We note that the outcome of the simulations does not change if we decrease the cell size to, e.g., $\unit[1]{nm} \times \unit[1]{nm} \times \unit[1]{nm}$.
\begin{figure}
	\centering
	\includegraphics[width=0.48\textwidth]{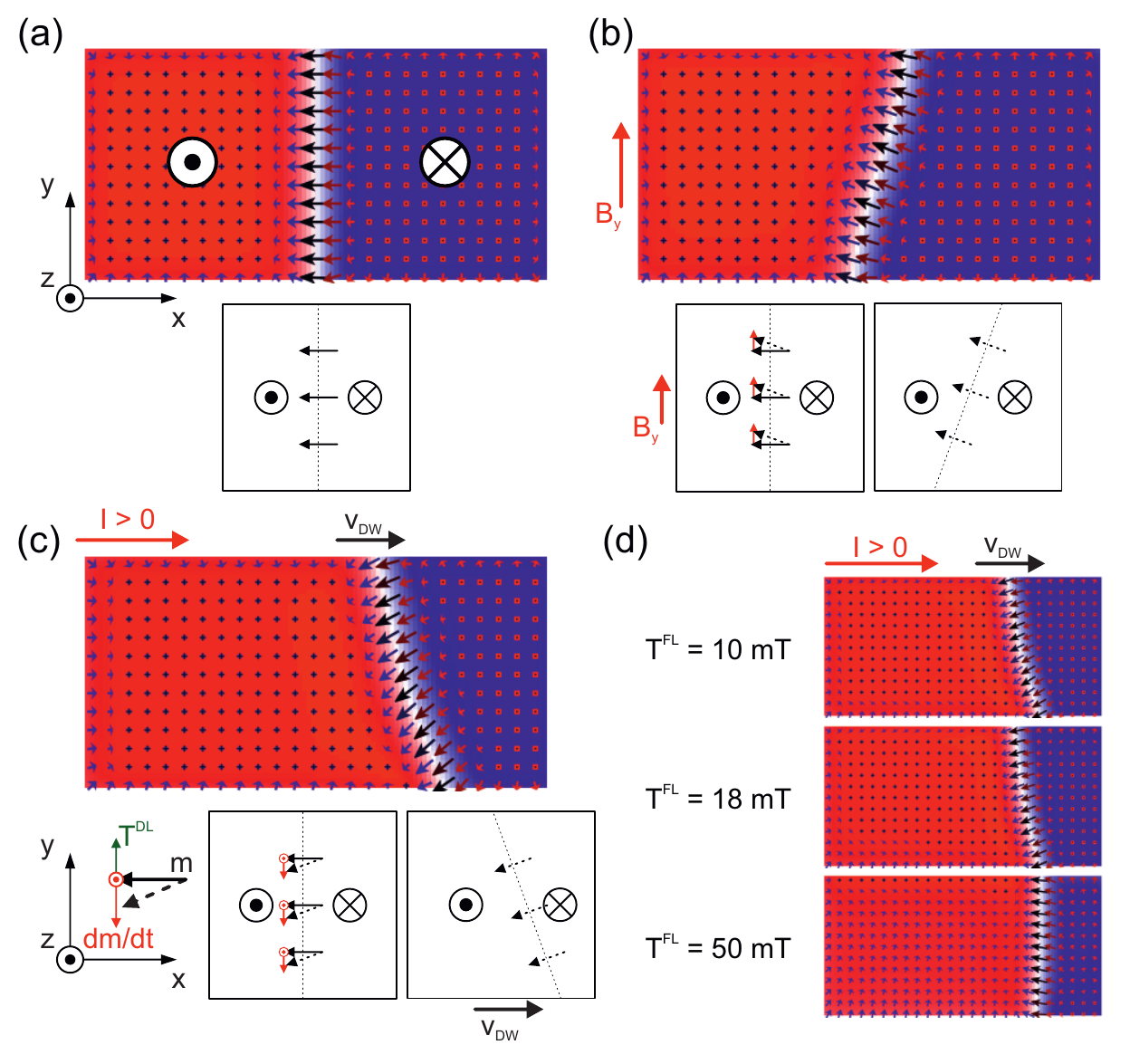}
	\caption{(a) Up/down DW in an AlO$_{\mathrm x}$/Co/Pt stripe at equilibrium. (b) Static DW tilt induced by a magnetic field $B_y = \unit[20]{mT}$. (c) Dynamic DW tilt due to $\mathbf{T}^{DL}$ during the injection of an electric current $j=\unit[2\times10^8]{A\ cm^{-2}}$. The schematics in (a-c) illustrate $d\mathbf{m}/dt$ according to Eq.~\eqref{LLG2:eqn} due to $\mathbf{T}^{DL}$ and the resulting DW tilt. (d) Dependence of the dynamic DW tilt on the amplitude of $T^{FL}$ relative to $T^{DL}=\unit[18]{mT}$ per $\unit[10^8]{A\ cm^{-2}}$ for the same current density as in (c).}
	\label{DW_tilt:pic}
\end{figure}
Figure~\ref{DW_tilt:pic}(a) shows the equilibrium configuration of an up/down DW in AlO$_{\mathrm x}$/Co/Pt, which is a left-handed N\'eel wall stabilized by the DMI. In order to illustrate the different mechanisms that lead to the tilting of the DW, we report in Figs.~\ref{DW_tilt:pic}(b) and (c) the response of such a DW to a transverse magnetic field $B_y$ and damping-like SOT $T^{DL}$, respectively. In Fig.~\ref{DW_tilt:pic}(b), $B_y$ rotates the DW moments away from the longitudinal direction towards $+\mathbf{y}$, which causes a negative tilt of the DW in order to maintain the energetically favoured N\'eel configuration. The equilibrium tilt is determined by the balance between external field, DMI, and DW energy, which increases with the DW length and hence with the tilt angle.\cite{Boulle:2013aa,Martinez:2014aa,Muratov:2017aa}
The effect of $T^{DL}$ due to a positive electric current (electrons flowing to the left) are shown in Fig.~\ref{DW_tilt:pic}(c). In order to understand the tilt of the DW in this case, we have to consider the action of the current-induced SOTs on the DW magnetization. The damping- and field-like SOTs have symmetry $\mathbf{T}^{DL}=T^{DL}\mathbf{m} \times \left( \mathbf{y}\times \mathbf{m}\right)$ and $\mathbf{T}^{FL}=T^{FL}\mathbf{m} \times \mathbf{y}$, respectively.\cite{Garello:2013aa}
The Landau-Lifshitz-Gilbert (LLG) equation is then given by
\begin{equation}
\frac{d\mathbf{m}}{dt}=-\frac{|\gamma|}{(1+{\alpha}^2)}\sum_{i}{\mathbf{T}_i} - \frac{|\gamma|\alpha}{(1+{\alpha}^2)}\mathbf{m}\times{\sum_{i}{\mathbf{T}_i}},
\label{LLG2:eqn}
\end{equation}
with
\begin{equation}
\sum_{i}\mathbf{T}_i = \mathbf{m}\times\mathbf{B}_{\mathrm{eff}} + \mathbf{T}^{DL} + \mathbf{T}^{FL}.
\end{equation}
Here, $\textbf{m}=\textbf{M}/M_s$ is the unit magnetization vector, $|\gamma|$ the electronic gyromagnetic ratio, $\mu_0$ the free space permeability and $\mathbf{B}_{\mathrm{eff}}= \mathbf{B}_{\mathrm{ext}}+\mathbf{B}_{\mathrm{K}}-\frac{1}{M_s}\frac{\delta E_{DMI}}{\delta \mathbf{m}}-\frac{1}{M_s}\frac{\delta E_{ex}}{\delta \mathbf{m}}$ 
the effective magnetic field. Here, $\mathbf{B}_{\mathrm{ext}}$ is the external magnetic field, $\mathbf{B}_{\mathrm{K}}=2K_u/M_s$ the effective out-of-plane anisotropy field (including the demagnetizing field), and the last two terms are the effective DMI and exchange magnetic fields. We consider first only the effect of the damping-like torque. In this case the LLG equation can be written in simplified form as $d\mathbf{m}/dt\propto-\mathbf{T}^{DL}-\alpha\mathbf{m}\times\mathbf{T}^{DL}$. Hence, the DW magnetization is deviated towards $-\mathbf{y}$ and $\mathbf{+z}$ by the damping-like torque, as shown schematically in Fig.~\ref{DW_tilt:pic}c. This dynamic process leads to the observed propagation (due to the $z$-component of $d\mathbf{m}/dt$) and tilting of the DW (due to the $y$-component of $d\mathbf{m}/dt$). A quantitative description of this process is given in terms of a one-dimensional model of DW propagation in Refs.~\onlinecite{Boulle:2013aa, Martinez:2014aa, Emori:2014ab}. The effect of the field-like torque can finally be understood in analogy with that of the magnetic field $B_y$, so that the DW tilt angle at steady state depends on the ratio $T^{FL}/T^{DL}$, as shown in Fig.~\ref{DW_tilt:pic}(d).

\begin{figure}
	\centering
	\includegraphics[width=0.48\textwidth]{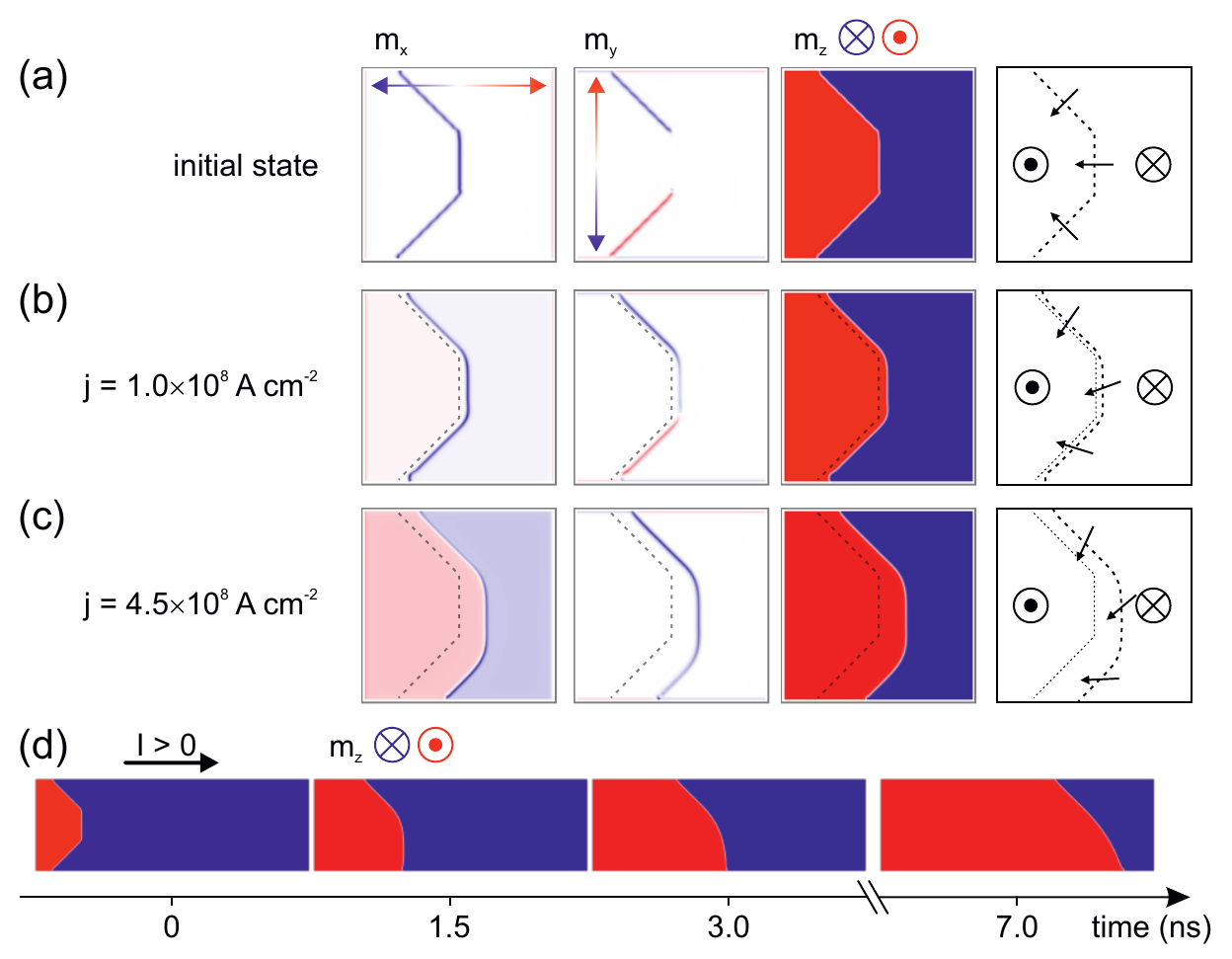}
	\caption{(a) Initial magnetic configuration of a Pt/Co/AlO\textsubscript{x} square with one straight and two oppositely tilted DWs. The side of the square is $\unit[1.5]{\mu m}$. The magnetization components $m_x$, $m_y$, and $m_z$ are shown in color in the different panels. The scheme on the right shows the in-plane magnetization and relative displacement of the DW. (b,c) Snapshot of the magnetic configuration during injection of a positive current of amplitude $j = \unit[1.0\times10^8]{A\ cm^{-2}}$ and $\unit[4.5\times10^8]{A\ cm^{-2}}$, respectively, taken after $\unit[0.9]{ns}$. The dotted lines show the initial DW position. (d) Snapshots of the DW propagation during injection of a positive current $j = \unit[4.5\times10^8]{A\ cm^{-2}}$ into a $\unit[4.5]{\mu m}$ long and $\unit[1.5]{\mu m}$ wide stripe. Note that after $\unit[\approx 1.5]{ns}$ the fastest DW is expelled from the stripe. As a consequence, the tilt angle at steady state corresponds to that of the slowest DW.}
	\label{ASYM-CIDWM_detail:pic}
\end{figure}
In order to investigate the relationship between the DW tilt angle $\psi$ and $v_{\mathrm{\scriptscriptstyle{DW}}}$, we simulate the dynamics of a DW consisting of one straight and two tilted sections in a square sample under the action of $\mathbf{T}^{DL}$ alone (Fig.~\ref{ASYM-CIDWM_detail:pic}). The magnetization on the left (right) side of the structure points along $+\mathbf{z}$ ($-\mathbf{z}$). We first relax the DW magnetization, which leads to the emergence of left-handed N\'eel walls. Due to the initial conditions, the three DWs have a tilt $\psi = \unit[-45]{^\circ}$, $\unit[0]{^\circ}$ and $\unit[45]{^\circ}$, shown in (a). Successive snapshots of the magnetic configuration during current injection reveal that the different DW components propagate with distinct velocities, as shown in Fig.~\ref{ASYM-CIDWM_detail:pic}(b,c). This behaviour can be easily understood in terms of Eq.~\eqref{LLG2:eqn} as $\mathbf{T}^{DL}$ 
rotates the DW magnetization against the effective DMI field towards $-\mathbf{y}$. As a result, for sufficiently large current, $m_x$ is largest (smallest) for $\psi = \unit[-45]{^{\circ}}$ ($\unit[45]{^{\circ}}$). Since $v_{\mathrm{\scriptscriptstyle{DW}}} \propto \left(d\mathbf{m}/dt\right)_z \propto {T}^{DL}$, and ${T}^{DL} \propto m_x$, $v_{\mathrm{\scriptscriptstyle{DW}}}$ is largest (smallest) for $\psi = \unit[-45]{^{\circ}}$ ($\unit[45]{^{\circ}}$). Therefore, the different $m_x$ components result in a pronounced asymmetry of the current-induced DW motion, as shown in (c). Alternatively, the difference in $v_{\mathrm{\scriptscriptstyle{DW}}}$ can be understood by an energy argument. Due to the presence of DMI, the energy is minimized if the DWs are of N\'eel type. During current injection, the DWs tilted at $\psi = \unit[0]{^{\circ}}$ and $ \unit[-45]{^{\circ}}$ deform and acquire a mixed N\'eel-Bloch character. These DWs propagate faster in order to reduce the total energy of the system by increasing the length of the energetically favoured N\'eel walls. The fastest direction of DW propagation measured by time-resolved scanning transmission x-ray microscopy\cite{Baumgartner:2017aa} as well as the largest displacements reported in "oblique" Pt/Co/AlO\textsubscript{x} racetracks oriented at different angles with respect to the current\cite{Safeer:2016aa} are consistent with this picture.

A relevant consequence of the asymmetric DW velocity is that, in an elongated stripe, the faster DWs ($\psi = \unit[0]{^{\circ}}$, $ \unit[-45]{^{\circ}}$) are rapidly expelled from the sample, and the final DW observed in steady state conditions is the slowest one with $\psi = \unit[45]{^{\circ}}$ [Fig.~\ref{ASYM-CIDWM_detail:pic}(d)]. This behavior has a compelling analogy with crystal growth, in which the crystal facets with the slowest growth rate determine the final crystal shape.\cite{Burton:1951} Similar arguments based on classical
interface thermodynamics explain the faceting observed during the growth of chiral magnetic bubbles subject to an applied field.\cite{Lau:2016ab} We thus conclude that the discrepancy between the DW configurations reported for quasi-static [Fig.~\ref{DW_tilt_symmetry:pic}(a)] and time-resolved measurements\cite{Ryu:2012aa,Baumgartner:2017aa} [Fig.~\ref{DW_tilt_symmetry:pic}(b)] is due to the different time-scales probed in these experiments, which correspond to the slower and faster DW in a racetrack, respectively. In time-resolved switching experiments, the initial conditions, namely the shape of the DW after nucleation, also play a role in determining the tilt and velocity of the DW. The final tilt angle is reached on a time scale of several ns, which increases with the stripe width.\cite{Boulle:2013aa}

\begin{figure}
	\centering
	\includegraphics[width=0.48\textwidth]{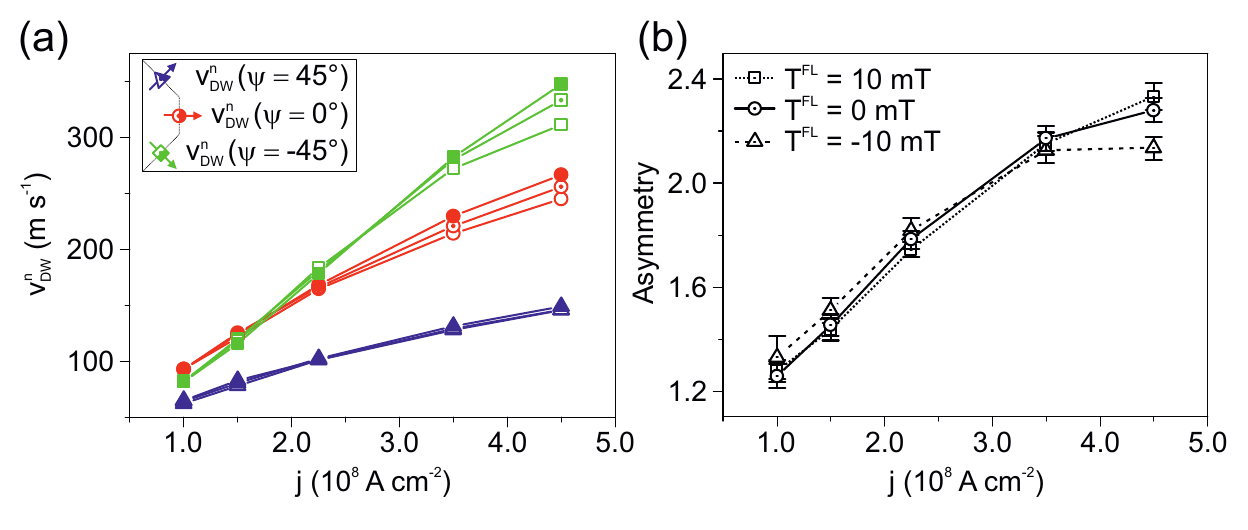}
	\caption{(a) Normal DW velocity $v_{\mathrm{\scriptscriptstyle{DW}}}^n$ as a function of current density for different tilt angles. The velocities are calculated for $T^{DL} = \unit[18]{mT}$ and $T^{FL} = \unit[10]{mT}$ (full symbols), $T^{FL} = \unit[0]{mT}$ (dotted symbols), and $T^{FL} = \unit[-10]{mT}$ (open symbols) per $j = \unit[10^8]{A\ cm^{-2}}$. (b) Asymmetry ratio $v_{\mathrm{\scriptscriptstyle{DW}}}^n(\psi = \unit[-45]{^{\circ}})/v_{\mathrm{\scriptscriptstyle{DW}}}^n(\psi = \unit[45]{^{\circ}})$, plotted as a function of current density for the three values of $T^{FL}$ shown in (a). Positive values of $T^{FL}$, as in Pt/Co/AlO\textsubscript{x}, reduce the asymmetry, whereas negative values increase it.}
	\label{ASYM-CIDWM_TFL:pic}
\end{figure}

The propagation velocity perpendicular to each DW front, $v_{\mathrm{\scriptscriptstyle{DW}}}^n(\psi = \unit[45]{^{\circ}})$, $v_{\mathrm{\scriptscriptstyle{DW}}}^n(\unit[\psi = 0]{^{\circ}})$, and $v_{\mathrm{\scriptscriptstyle{DW}}}^n(\unit[\psi = -45]{^{\circ}})$, can be calculated by measuring the distance travelled by the DW as a function of time. Figure~\ref{ASYM-CIDWM_TFL:pic}(a) shows that $v_{\mathrm{\scriptscriptstyle{DW}}}^n$ increases almost linearly with $j$ for all three DW components, however, with distinct slopes. Depending on $\psi$, $v_{\mathrm{\scriptscriptstyle{DW}}}^n$ for the fastest and slowest DW can differ by more than a factor two. Further, the asymmetry of $v_{\mathrm{\scriptscriptstyle{DW}}}^n$, which we define as the ratio $v_{\mathrm{\scriptscriptstyle{DW}}}^n(\unit[\psi = -45]{^{\circ}})/v_{\mathrm{\scriptscriptstyle{DW}}}^n(\unit[\psi = 45]{^{\circ}})$, increases proportionally to $j$ up to $\unit[3.5\times10^8]{A\ cm^{-2}}$, as shown in Fig.~\ref{ASYM-CIDWM_TFL:pic}(b).

Finally, we study the effect of the field-like torque on $v_{\mathrm{\scriptscriptstyle{DW}}}^n$. For a positive current, $\mathbf{T}^{FL}$ in Pt/Co/AlO\textsubscript{x} is equivalent to a magnetic field $B_y$ opposite to the Oersted field. Therefore, $\mathbf{T}^{FL}$ counteracts the rotation towards $-\mathbf{y}$ induced by the damping-like torque. More importantly, ${T}^{FL}>0~(<0)$ leads to an additional $\left( d\mathbf{m}/dt\right) _z$ contribution which increases (decreases) $v_{\mathrm{\scriptscriptstyle{DW}}}^n$. The amount of increase or decrease of $v_{\mathrm{\scriptscriptstyle{DW}}}^n$ due to the field-like torque depends on $\psi$, hence on the damping-like torque and DMI. We find that the ratio $v_{\mathrm{\scriptscriptstyle{DW}}}^n(T^{FL}>0)/v_{\mathrm{\scriptscriptstyle{DW}}}^n(T^{FL}<0)$ increases linearly as a function of $j$ for $\unit[\psi \neq  -45]{^{\circ}}$. Although the increase is only about 10\% at the highest $j$, this effect should not be neglected in devices with a significant field-like torque. These results are consistent with experiments in which an in-plane field $B_y$ was applied to reinforce the field-like torque, thus assisting the magnetization reversal\cite{Baumgartner:2017aa} and increasing the current-induced DW velocity.\cite{Emori:2013aa}

In summary, we reported a comparative study of the tilt and velocity of DWs in perpendicularly magnetized Pt/Co/AlO\textsubscript{x} layers. Consistently with qualitative arguments derived from the LLG equation, our micromagnetic simulations evidence that DWs with different tilt angles propagate at distinct speed, depending on the balance between DMI, damping-like, and field-like torque, which determines the $m_x$ component of the DW magnetization. As a result of the asymmetric speed of tilted DWs, the fastest DW in racetrack structures is expelled from the track after a time of the order of $\unit[1.5]{ns}$, which depends on the width of the track and initial shape of the DW. Thus, quasi-static measurements of the DW displacements induced by a sequence of current pulses probe the propagation and tilt of the slowest DW,\cite{Ryu:2012aa, Garg:2017aa, Boulle:2013aa, Martinez:2014aa} whereas time-resolved microscopy and "oblique" racetrack measurements probe the fastest DW.\cite{Safeer:2016aa, Baumgartner:2017aa} As a side remark, we note that the fastest propagation direction of the DW corresponds to the direction of motion of magnetic skyrmions, as described by the so-called "skyrmion Hall effect".\cite{Jiang:2017,Litzius:2017} Because a skyrmion is delimited by a DW with a tilt angle that varies continuously between $\unit[\psi = 0]{^{\circ}}$ and $\unit[\psi = 360]{^{\circ}}$ , the skyrmion Hall effect can be rationalized in terms of the preferential direction for DW propagation and the tendency of the skyrmions to retain their topologically protected shape.
These findings allow for a better understanding and tuning of the DW motion and switching speed of magnetic memory elements of different shapes.

\begin{acknowledgments}
We acknowledge funding by the Swiss National Science Foundation under grant 200020-172775. We acknowledge fruitful discussions with C. O. Avci.
\end{acknowledgments}

\end{document}